\documentclass[aip,jcp,showpacs,showkeys,reprint,eqsecnum,longbibliography]{revtex4-1}
\usepackage{amssymb}
\usepackage{amsmath}
\usepackage{amsbsy}
\usepackage{bm}
\usepackage{color}
\usepackage{graphicx}
\newcommand{\bq}{\begin{eqnarray}}
\newcommand{\eq}{\end{eqnarray}}
\newcommand{\bqn}{\begin{eqnarray*}}
\newcommand{\eqn}{\end{eqnarray*}}

\newcommand{\PP}[2]{\mbox{$(\phi_{#1},x_{#1}|\phi_{#2},x_{#2})$}}
\newcommand{\Li}{\text{Li}}

\begin{document}
%%%%%%%%%%%%%%%%%%%%%%%%%%%%%%%%%%%%%%%%%%%%%%%%%%%%%%%%%%%%%%%%%%%%%%%%%%%%%%
%%%%%%%%%%%%%%%%%%%%%%%%%%%%%%%%%%%%%%%%%%%%%%%%%%%%%%%%%%%%%%%%%%%%%%%%%%%%%%
%%%%%%%%%%%%%%%%%%%%%%%%%%%%%%%%%%%%%%%%%%%%%%%%%%%%%%%%%%%%%%%%%%%%%%%%%%%%%%
\title{Exact results for one dimensional fluids through functional
  integration} 

\author{Riccardo Fantoni}
\email{rfantoni@ts.infn.it}
\affiliation{Universit\`a di Trieste, Dipartimento di Fisica, strada
  Costiera 11, 34151 Grignano (Trieste), Italy}
%\affiliation{Dipartimento di Scienze Molecolari e Nanosistemi,
%  Universit\`a Ca' Foscari Venezia, Calle Larga S. Marta DD2137,
%  I-30123 Venezia, Italy} 

\date{\today}

\begin{abstract}
We review some of the exactly solvable one dimensional continuum fluid
models of equilibrium classical statistical mechanics under the
unified setting of functional integration in one dimension. We make
some further developments and remarks concerning fluids with
non pairwise-additive interaction. We then apply our developments to
the study of a particular non pairwise-additive Gaussian model for
which we are unable to find a well defined thermodynamics.
\end{abstract}



\maketitle
%%%%%%%%%%%%%%%%%%%%%%%%%%%%%%%%%%%%%%%%%%%%%%%%%%%%%%%%%%%%%%%%%%%%%%%%%%%%%%
\section{Introduction}
%%%%%%%%%%%%%%%%%%%%%%%%%%%%%%%%%%%%%%%%%%%%%%%%%%%%%%%%%%%%%%%%%%%%%%%%%%%%%%
\label{sec:introduction}

The physics of one dimensional systems is simpler than that one of 
higher dimensional ones. Specifically the free energy of an interacting
gas, a {\sl fluid}, has had an exact solution only in one dimension. The
apparent simplicity of restricting motion to one spatial dimension is
well known and has had much appeal. But what is the relation between
the exactly soluble models of the one dimensional world and the richer
and puzzling problems of the three dimensional one? A one dimensional
gas was once thought to be incapable even of condensation. Later with
the introduction of infinite range forces it has been made to condense,
but even so this liquid can never freeze. What one finds is that 
these models are useful tests of approximate mathematical methods,
the solutions of these models are surprisingly complex and
interesting, physical applications are often and unexpectedly
discovered, and more importantly they educate us to the need of
rigorous and exact analysis with which one can have a better
definition of reality. The fact that particles can get around each
other is responsible for much of the structure of the ordinary world,
and is also responsible for the difficulties which the mathematical
physicist encounter in studying it. In one dimension we renounce to
some of the structure in favor of the possibility of obtaining an
exact solution.

The importance of one dimensional physics also lies in the fact that
a number of many-body problems in higher dimensions can be accurately
mapped into one dimensional problems. 

L\'eon Van Hove showed that one dimensional fluids with impenetrable
particles each one interacting with a finite number of neighbors do
not have a phase transition at a non zero temperature
\cite{VanHove50}. 

In this work I will describe a way of simplifying the calculation
of the grand canonical partition function of an ensemble of classical
particles living in a one dimensional world and interacting with a
given pair-potential $v$, originally described by Edwards and Lenard
in their paper \onlinecite{Edwards62} which I will call EL from now
on. Using the notion of a general Gaussian random process and Kac's
theorem, they show how it is possible to express the grand partition
function as a one dimensional integral of the fundamental solution of
a given partial differential equation. The kind of partial
differential equation will be fixed by the kind of {\sl diffusion
  equation} satisfied by the Gaussian random process. In sections
\ref{sec:problem}, \ref{sec:averaging}, and \ref{sec:kac} I will
present EL's functional integration technique. In subsection
\ref{sec:wiener} I will show how, in EL, the properties of the Wiener
process are used to solve ``Edwards' model'' or ``Lenard's model''. I
will then show, in subsection \ref{sec:ornstein}, how one can use the
properties of the Ornstein-Uhlenbeck process to solve the
``Kac-Baker's model''. 

The main original contribution given in this work lies in section
\ref{sec:general} where I show how a generalized Ornstein-Uhlenbeck
process can be used to solve models with a more general non
pairwise-additive 
interaction potential. In Section \ref{sec:thermodynamics} I show how
EL propose to extract thermodynamical informations from their
functional integration treatment and in section
\ref{sec:characteristic} I show, following EL, how it is possible 
to reduce the search of the grand partition function, to a
characteristic value problem, when the diffusion equation is
independent of ``time''. In section \ref{sec:general} I show how one
has to renounce to this reduction, for the generalized
Ornstein-Uhlenbeck process satisfying, in general, a non separable
diffusion equation. In subsection \ref{sec:example} I then apply the
theoretical framework of such section to a non pairwise-additive
penetrable Gaussian interaction model. In 
particular I will prove that this model is thermodynamically unstable
in its attractive version (which is also not H-stable) and I will find
an approximate expression for the grand partition function of the
repulsive version (which clearly is H-stable) in terms of a triple
series one of which is alternating.   
   
More recently \cite{Dean98} the functional integral technique of
Edwards and Lenard has been used to solve the statistical mechanics of
a one dimensional Coulomb gas with boundary interactions as a one
dimensional model for a colloidal and soap film.

I think that the success of the functional integration method
described in this work to find exact solutions of the equilibrium
classical (non-quantum) statistical mechanics problem of one
dimensional fluids has certainly been one of the motivations for the
popularity acquired by functional integration after the pioneering
developments of Marc Kac and Richard Feynman. The link with the theory
of stochastic processes is just a beautiful example of how many
different theoretical frameworks come together in the few exact
solutions of classical many-body problems. 

%%%%%%%%%%%%%%%%%%%%%%%%%%%%%%%%%%%%%%%%%%%%%%%%%%%%%%%%%%%%%%%%%%%%%%%%%%%%%%
\section{The Problem}
%%%%%%%%%%%%%%%%%%%%%%%%%%%%%%%%%%%%%%%%%%%%%%%%%%%%%%%%%%%%%%%%%%%%%%%%%%%%%%
\label{sec:problem}

The problem is to simplify the calculation of the grand canonical
partition function of a system of particles in the segment $[0,L]$
whose positions are labeled by $x_i$ with $i=1,2,\ldots,N$, in thermal
equilibrium at a reduced temperature $\theta$, namely,
\bq \label{omega1}
\Omega=\sum_{N=0}^\infty \frac{z^N}{N!}\int_0^Ldx_N\cdots\int_0^Ldx_1
\,e^{-\frac{V_N(x1,\ldots,x_N)}{\theta}}.
\eq
EL consider the total potential energy of the system to be, 
\bq \label{potential}
V_N(x_1,\ldots,x_N)=\sum_{i=1}^N\sum_{j=1}^Nw(x_i,x_j),
\eq
where $w(x_i,x_j)$ is a function of two variables depending on the
pair-potential $v(|x_i-x_j|)$ and the kind of reservoir exchanging
particles with the system.

The main idea of EL, is to rewrite the grand partition function as a
functional average,
\bq \label{functional avarage}
\Omega &=&\left\langle e^{\int_0^Ldx^\prime\,F(\phi(x^\prime))}\right\rangle
\\ \nonumber
&=& \left\langle \sum_{N=0}^\infty \frac{1}{N!}\int_0^Ldx_N\cdots\int_0^Ldx_1
\,\prod_{i=1}^N F(\phi(x_i))\right\rangle.
\eq
And then choose $F(\phi)=z\exp(i\sigma\phi)$, to get,
\bq \nonumber
&&\Omega=\\ \label{omega2}
&&\sum_{N=0}^\infty \frac{z^N}{N!}\int_0^Ldx_N\cdots\int_0^Ldx_1\,
\left\langle e^{i\sigma\sum_{i=1}^N\phi(x_i)}\right\rangle,
\eq
where in interchanging the average with the sum and the integrals they
use the linearity of the average. we haven' t defined the average yet
so we will do it next. 

%%%%%%%%%%%%%%%%%%%%%%%%%%%%%%%%%%%%%%%%%%%%%%%%%%%%%%%%%%%%%%%%%%%%%%%%%%%%%%
\section{Averaging over a general Gaussian Random Process}
%%%%%%%%%%%%%%%%%%%%%%%%%%%%%%%%%%%%%%%%%%%%%%%%%%%%%%%%%%%%%%%%%%%%%%%%%%%%%%
\label{sec:averaging}

A general Gaussian random process $\phi(x)$ is defined by the postulate
that for any finite number of points $x_1,\ldots,x_N$ the joint 
probability density for $\phi(x_k)$ in $d\phi_k$ (we will often make use
of the abbreviation $\phi_i\equiv\phi(x_i)$) is of the form,
\bq \label{gaussian}
P(\phi_1,\ldots,\phi_N)=
\frac{\sqrt{\det B}}{(2\pi)^{N/2}} e^{ -\frac{1}{2}
\sum_{k=1}^N\sum_{l=1}^N B_{kl}\phi_k\phi_l },
\eq 
where $B_{ij}=B_{ij}(x_1,\ldots,x_N)$ are the elements of the positive
definite matrix $B$.

Let $\alpha_k$ be $N$ arbitrary real numbers. Then,
\bq \label{covariance1}
\left\langle e^{i\sum_{i=1}^N\alpha_i\phi_i}
\right\rangle&=&e^{ -\frac{1}{2}
\sum_{k=1}^N\sum_{l=1}^N C_{kl}\alpha_k\alpha_l },
\\ \nonumber
\mbox{where}~~~C&=&B^{-1}.
\eq
Differentiating with respect to $\alpha_k$ and $\alpha_l$ (not excluding
$k=l$) and then setting all $\alpha$ to zero, one obtains,
\bq \label{covariance2}
\langle\phi(x_k)\phi(x_l)\rangle=C_{kl}=C(x_k,x_l),
\eq
where $C$ is a function of two variables only, called the {\sl 
covariance function}. From equations (\ref{covariance1}) and 
(\ref{covariance2}) follows that also $B_{ij}=B(x_i,x_j)$ is
a function of just two variables. The covariance completely 
characterizes the statistical nature of $\phi(x)$

Replacing all the $\alpha$'s in equation (\ref{covariance1}) with
$\sigma$  and comparing (\ref{covariance1}) and (\ref{omega2}) with 
(\ref{omega1}) and (\ref{potential}) one recognizes that,
\bq \label{covariance-potential}
C(x_1,x_2)=\frac{2}{\theta \sigma^2}w(x_1,x_2).
\eq
This imposes a restriction to the systems that one can treat. Namely we
need $w$ to be positive definite.

Why is all this useful is explained in the next section.

\begin{widetext}
%%%%%%%%%%%%%%%%%%%%%%%%%%%%%%%%%%%%%%%%%%%%%%%%%%%%%%%%%%%%%%%%%%%%%%%%%%%%%%
\section{Kac's Theorem}
%%%%%%%%%%%%%%%%%%%%%%%%%%%%%%%%%%%%%%%%%%%%%%%%%%%%%%%%%%%%%%%%%%%%%%%%%%%%%%
\label{sec:kac}

Consider a Markoffian process $\phi(x)$, i.e. one for which, given any
increasing sequence of ``times'' $x_0,x_1,\ldots,x_n$, with $x_0\le
x_1\le\cdots\le x_n$, the probability density that $\phi(x_k)$ is in
$d\phi_k$ (with $k=0,1,\ldots,n$) is the product, 
\bq \label{wiener}
P(\phi_1,\ldots,\phi_n)=\int_{-\infty}^{\infty}
\prod_{k=1}^nP\PP{k}{k-1}R(\phi_0,x_0)d\phi_0,
\eq   
where $P\PP10$ is the conditional probability that $\phi(x_1)$ is in an 
element $d\phi_1$ around $\phi_1$ given that $\phi(x_0)=\phi_0$  and 
$R(\phi,x)$ is the initial probability distribution for the process.
\footnote{Equation (\ref{wiener}) defines what is often called a 
{\sl Wiener measure} in the space of continuous functions $\phi(x)$.}
Both the conditional probabilities and the initial distribution are 
assumed to be normalized to unity over the interval $\phi\in[-\infty
,+\infty]$, 
\bq \label{normalization}
\int_{-\infty}^{\infty} d\phi_1\,P\PP10=\int_{-\infty}^{\infty} d\phi\,
R(\phi,x)=1.
\eq
Any quantity which is an expression involving $\phi(x)$ is a random
variable whose average value may be determined using the probability
(\ref{wiener}).

One is interested in averages of the form,
\bq 
W(x,x_0)&=&\left\langle e^{ \int_{x_0}^xdx^\prime F(\phi(x^\prime),x^\prime)}
\right\rangle\\ \nonumber
&=&1+\sum_{n=1}^\infty\frac{1}{n!}\int_{x_0}^xdx_n\int_{x_0}^{x}dx_{n-1}\cdots
\int_{x_0}^{x}dx_1\,\langle F(\phi_n,x_n)\cdots F(\phi_1,x_1)\rangle
\\ \nonumber
&=&1+\sum_{n=1}^\infty\int_{x_0}^xdx_n\int_{x_0}^{x_n}dx_{n-1}\cdots
\int_{x_0}^{x_2}dx_1\,\langle F(\phi_n,x_n)\cdots F(\phi_1,x_1)\rangle.
\eq
Kac's theorem takes advantage of the Markoffian property (\ref{wiener})
to relate to each other the successive terms of this series by an 
integral-recursion formula. It can be seen by inspection that,
\bq
W(x,x_0)&=&\int_{-\infty}^\infty d\phi\,Q\PP{}{0},\\ \nonumber
Q&=&\sum_{n=0}^\infty Q_n,\\ \nonumber
&&\left\{\begin{array}{l} \displaystyle
Q_0\PP{}{0}=\int_{-\infty}^\infty d\phi_0\,P\PP{}{0}R(\phi_0,x_0)\\
\displaystyle
Q_n\PP{}{0}=\int_{x_0}^xdx^\prime\int_{-\infty}^\infty d\phi^\prime\,
P(\phi,x|\phi^\prime,x^\prime)F(\phi^\prime,x^\prime)
Q_{n-1}(\phi^\prime,x^\prime|\phi_0,x_0)
\end{array}\right.
\eq
Then one can write the following integral equation for $Q$,
\bq \nonumber
Q\PP{}{0}&=&Q_0+\sum_{n=1}^\infty Q_n = \int d\phi_0\,PR+
\sum_{n=1}^\infty\int dx^\prime \int d\phi^\prime\,PFQ_{n-1}\\ \label{kac}
&=&\int_{-\infty}^\infty d\phi_0\,P\PP{}{0}R(\phi_0,x_0)
+\int_{x_0}^xdx^\prime\int_{-\infty}^\infty d\phi^\prime\,
P(\phi,x|\phi^\prime,x^\prime)F(\phi^\prime,x^\prime)
Q(\phi^\prime,x^\prime|\phi_0,x_0).
\eq
This is the main result of Kac's theorem.
\end{widetext}

Now assuming that the stochastic process $\phi(x)$ satisfies a forward 
Fokker-Planck equation,
\bq \label{fokker}
\frac{\partial}{\partial x}P\PP{}{0}&=&{\cal L}(\phi,x)P\PP{}{0}
\\ \nonumber
P(\phi,x_0|\phi_0,x_0)&=&\delta(\phi-\phi_0)~~~\mbox{initial condition}
\eq
it immediately follows from the integral formula (\ref{kac}), that
$Q$ satisfies,
\bq \nonumber
\frac{\partial}{\partial x}Q\PP{}{0}&=&[{\cal
    L}(\phi,x)+F(\phi,x)]\times \\ \label{kac theorem}
&&Q\PP{}{0}
\\ \nonumber
Q(\phi,x_0|\phi_0,x_0)&=&R(\phi,x_0)~~~~\mbox{initial condition}
\eq

If we now further assume $\phi(x)$ to be a Gaussian process (so that
equation (\ref{wiener}) is of the form (\ref{gaussian})) then we can 
put together the result of the previous section
(\ref{covariance-potential}) and Kac's theorem, to say that,
\bq \label{gcpf0}
\mbox{\fbox{$\displaystyle\Omega=W(L,0)=\int_{-\infty}^\infty d\phi\, 
Q(\phi,L|0,0)$}}, 
\eq
{\sl where $Q=Q\PP{}{0}$ is the solution of the partial differential 
equation (\ref{kac theorem}) with} $F(\phi,x)$
$=F(\phi)=z\exp(i\sigma\phi)$. This is the simplification found by EL.

Note the following points:
\begin{itemize}
\item This certainly is a simplification from a computational point of 
view and establishes a bridge between non-equilibrium statistical
mechanics and the theory of stochastic processes and equilibrium
statistical mechanics in one dimension. 
\item When the operator ${\cal L}$ is independent of ``time'' (we keep
calling $x$ time because it comes natural from the notion of random 
process. In the present context though $x$ is the position of a particle
along his one dimensional world) then both $P\PP{}{0}$ and $Q\PP{}{0}$  
depend only on $|x-x_0|$ since $F$ does not depend explicitly on $x$. 
\item For a non-stationary random process $\phi(x)$ it is often possible
to choose a delta function as initial distribution, i.e. $R(\phi,x_0)=
\delta(\phi-\phi_0)$, where $\phi_0=\phi(x_0)$. In this case $Q$ is the 
{\sl fundamental solution} of the partial differential equation 
(\ref{kac theorem}).
\item For a non-stationary random process the covariance function 
$C(x_1,x_2)=\langle\phi(x_1)\phi(x_2)\rangle$ is not a function of 
$|x_2-x_1|$ alone. The identification of the covariance with the
pair-potential $v$ demands that the process be stationary because 
the pair-potential is a function of the difference of the two 
position variables. But in some cases (due for example to the presence 
of the reservoir) $w$ may differ from $v$ (see subsection
\ref{sec:wiener}). 
\end{itemize}
As a final remark, in EL is stressed the importance of the Markoffian
nature of the process. They observe that the concept of a Markoffian
process involves the idea of a succession in ``time'' and this is
meaningless when there is more then one independent variable. So it
seems to be hard to extend the technique just described even to a two
dimensional system.

In the following section we will apply the functional integration
technique just described to some concrete example.

%%%%%%%%%%%%%%%%%%%%%%%%%%%%%%%%%%%%%%%%%%%%%%%%%%%%%%%%%%%%%%%%%%%%%%%%%%%%%%
\section{Examples}
%%%%%%%%%%%%%%%%%%%%%%%%%%%%%%%%%%%%%%%%%%%%%%%%%%%%%%%%%%%%%%%%%%%%%%%%%%%%%%
\label{sec:examples}

Note that due to the Markoffian nature of the stochastic process the
following two properties should be required for $x_0\le x_1\le x_2$, 
\bq \label{MP1}
&&R(\phi_1,x_1)=\int_{-\infty}^\infty d\phi_0\,P\PP{1}{0} R(\phi_0,x_0),\\ \nonumber 
&&P\PP{2}{0}=\\ \label{MP2}
&&\int_{-\infty}^\infty d\phi_1\,P\PP{2}{1}P\PP{1}{0}.
\eq

Let us see now how all this works for two well known Markoffian,
Gaussian stochastic processes.

%%%%%%%%%%%%%%%%%%%%%%%%%%%%%%%%%%%%%%%%%%%%%%%%%%%%%%%%%%%%%%%%%%%%%%%%%%%%%%
\subsection{The Ornstein-Uhlenbeck process}
%%%%%%%%%%%%%%%%%%%%%%%%%%%%%%%%%%%%%%%%%%%%%%%%%%%%%%%%%%%%%%%%%%%%%%%%%%%%%%
\label{sec:ornstein}

The {\sl Ornstein-Uhlenbeck process} is a stationary process defined
as follows,
\bq
R(\phi_0,x_0)&=&\frac{e^{-\frac{\phi_0^2}{2}}}{\sqrt{2\pi}},\\
P\PP{}{0}&=&\frac{e^{-\frac{\left(\phi-\phi_0e^{-\gamma \Delta
        x}\right)^2}{2S(\Delta x)}}}{\sqrt{2\pi S(\Delta x)}}, 
\\ \nonumber
\mbox{with}~~~~\Delta x&=&|x-x_0|,\\ \nonumber
S(\Delta x)&=&1-e^{-2\gamma \Delta x},
\eq
where $\gamma$ is the inverse of the characteristic time constant of
the process, i.e. a positive real number. 

The covariance for this process is,
\bq
C(x_1,x_2)=e^{-\gamma|x_1-x_2|}.
\eq

The Fokker-Planck equation satisfied by the process is the Smoluchowski
diffusion equation for an harmonic oscillator,
\bq
{\cal L}(\phi)=\gamma\left(\frac{\partial^2}{\partial\phi^2}
+\frac{\partial}{\partial\phi}\phi\right).
\eq

So this process can be used to describe a system of particles whose
potential energy is,
\bq \label{kac-baker}
w(x_1,x_2)=\frac{\theta\sigma^2}{2}e^{-\gamma|x_1-x_2|}.
\eq
Adding a hard-core part to this long range potential and making
it attractive by choosing $\sigma$ pure imaginary, gives the so called
``Kac-Baker model''. Yang and Lee showed that the presence of the hard 
core part is sufficient to ensure the existence of the thermodynamic 
potential for the infinite system ($L\rightarrow \infty$). 
This was calculated exactly by Kac 
who also proved that the model has no phase transitions (because of
the infinite range of the potential, L. Van Hove's proof is not applicable
here). Later Baker showed that if one sets,
\bq
\sigma=i\sqrt{\frac{\alpha_0\gamma}{\theta}},
\eq
(so that the integral of the potential is independent of $\gamma$) and
then takes the limit $\gamma\rightarrow 0$ {\sl after} the limit
$L\rightarrow \infty$, {\sl then a phase transition of the classical 
Van der Waals type is obtained.} A model with exponential repulsive 
pair-potential (exactly like the one in (\ref{kac-baker})) was studied
by D. S. Newman, who concluded that it did not show phase transitions 
in the long range limit $\gamma\to 0$. \cite{Lieb66} 
 
%%%%%%%%%%%%%%%%%%%%%%%%%%%%%%%%%%%%%%%%%%%%%%%%%%%%%%%%%%%%%%%%%%%%%%%%%%%%%%
\subsection{The Wiener process}
%%%%%%%%%%%%%%%%%%%%%%%%%%%%%%%%%%%%%%%%%%%%%%%%%%%%%%%%%%%%%%%%%%%%%%%%%%%%%%
\label{sec:wiener}

We follow EL and introduce the {\sl Wiener process}. It is a
non-stationary process defined by (if $x\ge x_0>0$),
\bq
R(\phi_0,x_0)&=&\frac{e^{-\frac{\phi_0^2}{4Dx_0}}}{\sqrt{4\pi D x_0}}\\ 
P\PP{}{0}&=&\frac{e^{ -\frac{\Delta \phi^2}{4 D 
\Delta x}}}{\sqrt{4\pi D\Delta x}}, \\ \nonumber
\mbox{with}~~~~\Delta x&=&x-x_0,\\ \nonumber
\Delta \phi&=&\phi-\phi_0,
\eq
where $D$ is the diffusion constant of the Brownian process, i.e. a 
positive real number.

The covariance for this process is,
\bq
C(x_1,x_2)=2D\min(x_1,x_2).
\eq
Although this process cannot be differentiated it can be seen as the
integral, $\phi(x)=\int_0^xds\,\xi(s)$, of the Gaussian white noise
process, $\xi(x)$, defined by $\langle\xi(x)\rangle=0$
and $\langle\xi(x_1)\xi(x_0)\rangle=\zeta^2\delta(x_1-x_0)$ and the
attribute Gaussian implies that all cumulants higher than of second
order vanish. One just needs to set $2D=\zeta^2$.

The Fokker-Planck equation satisfied by the process is the Einstein
diffusion equation,
\bq
{\cal L}(\phi)=D\frac{\partial^2}{\partial\phi^2}.
\eq

So this process can be used to describe a system of particles whose
potential energy is,
\bq
w(x_1,x_2)=D\theta\sigma^2\min(x_1,x_2).
\eq
It was S. F. Edwards, see EL,  who first realized that this is
a Coulomb system:
electrons of charge $q$ living in the segment $[0,L]$ are in contact
with an infinite reservoir (in the region $x<0$, say). The reservoir
exchanges particles with the system of electrons giving rise to the 
statistical fluctuations in particle number. Take the system plus 
reservoir electrically neutral as a whole and imagine the system 
containing $N$ electrons. Then there is a total charge $-Nq$ in the 
reservoir. Gauss theorem then tells that in the region $x\ge 0$
there is a constant electric field of magnitude $2\pi N q$, due to 
the presence of the reservoir. Now choosing,
\bq
D&=&\frac{2\pi}{\theta},\\
\sigma &=&q,
\eq
one can rewrite the total potential energy of the system as,
\bq \nonumber
V_N&=&2\pi q^2\sum_{k=1}^N\sum_{l=1}^N\min(x_k,x_l)\\ \nonumber
&=&2\pi q^2\sum_{k=1}^N\sum_{l=1}^N\left[
-\frac{|x_k-x_l|}{2}+\frac{x_k+x_l}{2}\right]\\ \nonumber
&=&-2\pi q^2\sum_{k<l}|x_k-x_l|+2\pi q^2\sum_{k=1}^N\sum_{l=1}^Nx_l\\
&=&-2\pi q^2\sum_{k<l}|x_k-x_l|+2\pi N q^2\sum_{l=1}^Nx_l.
\eq
Which is readily recognized as the expected result for the ``Edwards'
model''. We are assuming that the line is the real world in which 
each charge lives. So that also its field lines cannot escape from
the line. Then the electric potential of each charge is the 
solution of $d^2\psi(x)/dx^2=-4\pi\delta(x)$, i.e. $\psi(x)=-2\pi|x|$.

Note that due to the presence of the neutralizing reservoir, $w$ is not 
just a function of $|x_i-x_j|$ and consequently the random process is not 
just a stationary one as in the Kac-Baker example.

In this case Edwards has not been able to answer in a definite way 
to the problem of continuity of the thermodynamic functions.

%%%%%%%%%%%%%%%%%%%%%%%%%%%%%%%%%%%%%%%%%%%%%%%%%%%%%%%%%%%%%%%%%%%%%%%%%%%%%%
\section{Thermodynamics}
%%%%%%%%%%%%%%%%%%%%%%%%%%%%%%%%%%%%%%%%%%%%%%%%%%%%%%%%%%%%%%%%%%%%%%%%%%%%%%
\label{sec:thermodynamics}

Following EL, we want now comment briefly on the relevance
of all this from the point of view of the thermodynamics of the system
of particles. Given the grand canonical partition function $\Omega=\Omega
(z,L,\theta)$ the equation of state follows from eliminating $z$ between
the two following equations,
\bq
\frac{P}{\theta}&=&\frac{1}{L}\ln\Omega(z,L,\theta),\\
n&=&z\frac{\partial}{\partial z}\frac{1}{L}\ln\Omega(z,L,\theta).
\eq
where $P$ is the pressure and $n$ the number density of particles.
Sometimes one talks about {\sl chemical potential} $\mu$ (of the 
one-component system), instead of $z$. The two are related by,
\bq
z=\left(\frac{m \theta}{2\pi\hbar^2}\right)^{1/2}e^{\mu/\theta} > 0,
\eq
where $m$ is the mass of the particles. All the other thermodynamic 
functions can be obtained from the internal energy,
\bq
U(N,L,S)&=&-\frac{\partial}{\partial(1/\theta)} \ln\Omega(z,L,\theta)
+\frac{1}{2}N\theta,
\eq
where $S$ is the entropy of the system. Or alternatively from the 
Helmholtz free energy,
\bq
A(N,L,\theta)=\mu N-\theta\ln\Omega(z,L,\theta).
\eq

It is often useful to simplify the problem by studying just the asymptotic 
behavior of $\Omega$ in the infinite system limit $L\rightarrow \infty$.
This usually allows the recognition of eventual phase transitions 
(as in the Yang and Lee theory and L. Van Hove theorem) as singularities
in the equation of state. The equation of state for the infinite system 
becomes then,
\bq
\mbox{\fbox{$
\left\{
\begin{array}{l} \displaystyle
\frac{P}{\theta}=\Phi(z,v,\theta)=\lim_{L\rightarrow\infty}
\left[\frac{1}{L}\ln\Omega(z,L,\theta)\right],\\  \displaystyle
n=\frac{1}{v}=\lim_{L\rightarrow\infty}\left[z\frac{\partial}
{\partial z}\frac{1}{L}\ln\Omega(z,L,\theta)\right],
\end{array}
\right.$}}
\eq
where the limit may not be freely interchanged with the differentiation.

%%%%%%%%%%%%%%%%%%%%%%%%%%%%%%%%%%%%%%%%%%%%%%%%%%%%%%%%%%%%%%%%%%%%%%%%%%%%%%
\section{Characteristic value problem}
%%%%%%%%%%%%%%%%%%%%%%%%%%%%%%%%%%%%%%%%%%%%%%%%%%%%%%%%%%%%%%%%%%%%%%%%%%%%%%
\label{sec:characteristic}

Both the examples described
have the common feature that ${\cal L}$ is independent of time $x$. 
Under this circumstance the problem of calculating the grand canonical
partition function $\Omega$ may be simplified even further, as shown
in EL.

Letting $\phi\to\phi/\sigma$, the coefficient function $F(\phi)$ in
equation (\ref{kac theorem}) is periodic with period $2\pi$. It is
then possible to reduce the problem (\ref{kac theorem}) to the
characteristic value problem of an ordinary differential operator on a
finite interval of the independent variable $\phi$. Let,
\bq
\tilde{Q}(\phi,x)=\sum_{n=-\infty}^\infty Q(\phi+2\pi n,x|0,0).
\eq
This function is the {\sl periodic solution} of the partial
differential equation (\ref{kac theorem}) and for $x=0$ 
it reduces to,
\bq
\tilde{Q}(\phi,0)=\sum_{n=-\infty}^\infty R(\phi+2\pi n,0).
\eq
For the ``Kac-Baker model'' one finds for example 
$\tilde{Q}(\phi,0)=\theta_3\left(i\pi\phi/\sigma^2,e^{-2\pi^2/\sigma^2}\right)
e^{-\phi^2/2\sigma^2}/\sqrt{2\pi\sigma^2}$, 
where $\theta_3$ is an elliptical theta function \cite{Abramowitz}, and
for the ``Edwards' model''
$\tilde{Q}(\phi,0)=\sum_{n=-\infty}^\infty\delta(\phi+2\pi n)$. So, for
this latter case, $\tilde{Q}$ is the {\sl periodic fundamental
  solution} of (\ref{kac theorem}). It then follows that, 
\bq \label{gcpf}
\Omega=\int_{-\pi}^\pi d\phi\, \tilde{Q}(\phi,L).
\eq
Since $F$ and ${\cal L}$ do not depend on $x$, in solving (\ref{kac theorem})
for $\tilde{Q}$, one may use the method of separation of variables. 
This leads to the characteristic value problem,
\bq
\left[{\cal L}(\phi)+F(\phi)\right]y(\phi)&=&\lambda y(\phi),
\\ \nonumber
y(\phi+2\pi)&=&y(\phi).
\eq
Then one looks for a complete orthonormal set of eigenfunctions $y_m$ 
with relative eigenvalues $\lambda_m$ ($m=0,1,2,\ldots$),
\bq
\int_{-\pi}^\pi d\phi\, y_m(\phi)y_{m^\prime}(\phi)=\delta_{m,m^\prime}.
\eq 
The expansion of $\tilde{Q}$ in terms of these functions is,
\bq
\tilde{Q}(\phi,x)&=&\sum_{m=0}^\infty e^{\lambda_m x}B_my_m(\phi),\\
B_m&=&\int_{-\pi}^\pi d\phi\,\tilde{Q}(\phi,0) y_m(\phi).
\eq
For example $B_m=y_m(0)$ for the ``Edwards' model''. The grand
partition function becomes, 
\bq
\Omega(L)&=&\sum_{m=0}^\infty A_m e^{\lambda_m L},\\
A_m&=&B_m\int_{-\pi}^\pi d\phi\,y_m(\phi).
\eq
The $\lambda_m$ and the $y_m$ depends parametrically on $z$ which 
enters into the definition of $F(\phi)$. Moreover since
$F(\phi)=F^*(-\phi)$ the $\lambda_m$ are either real or occur in
complex conjugate pairs.

Now assume that among the sequence of eigenvalue $\lambda_m$ there 
is one $\lambda_0$ {\sl that is real and is bigger than the real part 
of all the others} then the following simplification holds,
\bq
\mbox{\fbox{$
\Omega(L\rightarrow \infty) \sim A_0e^{\lambda_0 L}$}}.
\eq
The equation of state for the infinite system then becomes,
\bq
P&=&\theta\lambda_0(z),\\ \nonumber
n&=&\lim_{L\rightarrow \infty} \left[z\frac{\partial}
{\partial z}\left(\frac{\ln A_0(z)}{L}+\lambda_0(z)\right)\right]\\
&=&z\frac{\partial}{\partial z}\lambda_0(z).
\eq
For example for the ideal gas, $\sigma\to 0$ and $\lambda_0(z)=az$,
with $a$ a constant. 

Let us summarize the characteristic value problem for the examples
described. Denoting with a dash a first derivative respect to $\phi$
(${\ldots}^\prime\equiv d\ldots/d\phi$) we have:\\ 
(i) ``Kac-Baker model'' repulsive \cite{Lieb66},
\bq \label{KBM}
\gamma[\sigma^2 y^{\prime\prime}+(\phi y)^\prime]+ze^{i\phi}y=\lambda y,
\eq
(ii) ``Edwards' model'' \cite{Edwards62},
\bq \label{EM}
\frac{2\pi q^2}{\theta}y^{\prime\prime}+ze^{i\phi}y=\lambda y,
\eq
this is the one component plasma or {\sl jellium} system.\\
(iii) ``Lenard's model'' \cite{Lenard61},
\bq \label{LM}
\frac{2\pi q^2}{\theta}y^{\prime\prime}+2z\cos(\phi) y=\lambda y,
\eq
this is the two component plasma system of two kinds of particles with
charges $\pm q$ and the corresponding values of $z$ that by symmetry
may be assumed equal without loss of generality.

In all cases $y(\phi)$ is a function of period $2\pi$ (for 
the attractive Kac-Baker model the periodicity is lost but the 
characteristic value problem is still valid). 

Unfortunately there is no simple way to solve explicitly
Eq. (\ref{KBM}) for the Kac-Baker model. Nonetheless it is apparent
the existence of the thermodynamic limit for the repulsive model, as
was proved by D. S. Newman \cite{Newman1964}. 

In the Edwards' model the presence of the neutralizing reservoir is
responsible (the potential energy of interaction between the particles
and the reservoir being proportional to $+x$) for the charges all of
the same sign to accumulate at the 
origin resulting in  a system with zero density and pressure in accord
with the fact that Eq. (\ref{EM}) admits solutions in terms of
modified Bessel functions of the first kind 
$I_{\pm i\sqrt{2\theta\lambda/\pi q^2}}(\sqrt{2\theta z e^{i\phi}/\pi
q^2})$ which form a complete set for $\lambda=-m^2$ with
$m=0,1,2,\ldots$, so that $\lambda_0=0$.

\begin{figure}[htbp]
\begin{center}
\includegraphics[width=8cm]{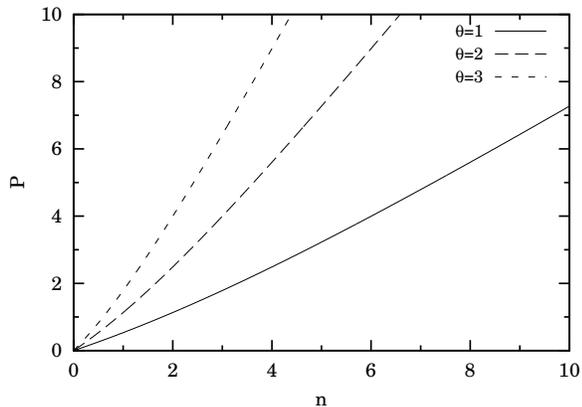}
\end{center}  
\caption{The equation of state for the Lenard's model
  at various temperatures $\theta$ for $q=1$.}
\label{fig:LM}
\end{figure}

In the Lenard's model the solutions of Eq. (\ref{LM}) is in terms of
even and odd Mathieu functions with characteristic value
$a=-2\lambda\theta/\pi q^2$, parameter $q=-2\theta z/\pi q^2$, and
argument $\phi/2$. According to Floquet's theorem, any Mathieu
function of argument $\phi$ can be written in the form
$e^{ir\phi}f(\phi)$, where $f$ has period $2\pi$ and $r$ is the {\sl
  Mathieu characteristic exponent}. For nonzero $q$ the Mathieu
functions are only periodic for certain values of $a$. Such {\sl
  Mathieu characteristic values} are given by $a_r=A(r,q)$ with $r$
integer or rational and $A(0,q)\le A(r,q)$ for all $r,q$. Then we will
have $\lambda_0=-(\pi q^2/2\theta)A(0,-2\theta z/\pi q^2)$. In
Fig. \ref{fig:LM} we show the equation of state of the Lenard model at
various temperatures $\theta$ for $q=1$.  
We are then led to conclude that this system does not admit any
phase transition, condensation (gas-liquid) or freezing
(liquid-solid).  

%%%%%%%%%%%%%%%%%%%%%%%%%%%%%%%%%%%%%%%%%%%%%%%%%%%%%%%%%%%%%%%%%%%%%%%%%%%%%%
\section{A non pairwise-additive penetrable interaction model}
%%%%%%%%%%%%%%%%%%%%%%%%%%%%%%%%%%%%%%%%%%%%%%%%%%%%%%%%%%%%%%%%%%%%%%%%%%%%%%
\label{sec:general}

In the examples described we started from known stochastic processes
to find which physical fluid model they may be able to describe. Actually
one wants to do the reverse: given a physical model, i.e. given $w$ (a
positive definite function (\ref{covariance-potential})), determine
the stochastic process that allows the desired simplification for the
grand canonical partition function. 

In the more general case one has to deal with $w$'s which are not
functions of the pair-potential alone, as happened in the case of
Edwards' model. For example one may be interested in modifying
Edwards' model for the case of a Coulomb system moving but not living
in $[0,L]$ with field lines allowed to exit the segment and
interacting with the full three dimensional pair-potential
$v(x)=1/\sqrt{x^2+\varepsilon^2}$, with $\varepsilon$ a small
positive quantity so that $v_0=1/\epsilon$ or
$\sigma=2/\sqrt{\varepsilon\theta}$. A neutralizing uniform
background in this case gives rise to quadratic terms making even the
one-component system stable. To obtain the purely one dimensional case
it is necessary to take the $\varepsilon\to 0$ limit at the end of the
analysis of the {\sl quasi one dimensional} case. 
This problem has been solved by R. J. Baxter \cite{Lieb66} who
developed a method for finding the partition function when the
pair-potential satisfies a linear differential equation with constant  
coefficients. His method still leads to an eigenvalue problem but does 
not employ functional averaging.

As a progress in this direction it is useful to reconsider the
Ornstein-Uhlenbeck process in a more general way. Consider the
following stationary stochastic process,
\bq \label{GOU1}
R(\phi_0,x_0)&=&\frac{e^{-\frac{\phi_0^2}{2}}}{\sqrt{2\pi}},\\ \label{GOU2}
P\PP{}{0}&=&\frac{e^{
-\frac{\left(\phi-\phi_0 A(\Delta x)\right)^2}{2S(\Delta
  x)}}}{\sqrt{2\pi S(\Delta x)}}, \\ \nonumber
\mbox{with}~~~~\Delta x&=&|x-x_0|,\\ \nonumber
S(\Delta x)&=&1-A^2(\Delta x),
\eq
where the last definition assures the validity of the Markoffian
property (\ref{MP1}). Clearly, in order to satisfy the Markoffian
property (\ref{MP2}) we need to require $A(x)A(y)=A(x+y)$ which is only
satisfied by choosing $A$ as an exponential as in the
Ornstein-Uhlenbeck process. Here we willingly violate this second
property and choose $A$ as an arbitrary function. In order to have 
$P(\phi,x_0|\phi_0,x_0)=\delta(\phi-\phi_0)$ we must also require that
$\lim_{x\to 0}A(x)=1$.

The covariance for this process is,
\bq
C(x_i,x_j)=\frac{2}{\theta\sigma^2}w(x_i,x_j)=
\prod_{k=i}^{j-1}A(|x_{k}-x_{k+1}|),
\eq
with $x_i\le x_{i+1}\le x_{i+2}\le\ldots \le x_j$. So the interaction
between particle $i$ and particle $j$ depends on how many particles lie
between them. This is a particular {\sl non pairwise-additive}
interaction model. 

It can be readily verified that the transition density of this process
satisfies the following forward Fokker-Planck equation,
\bq
{\cal L}(\phi,x)=-\frac{\dot{A}}{A}\left(\frac{\partial^2}{\partial\phi^2}
+\frac{\partial}{\partial\phi}\phi\right),
\eq
where the dot denotes differentiation with respect to time
($\dot{\ldots}\equiv d\ldots/dx$). All the properties of section
\ref{sec:kac} continue to hold. Then, calling $v=A\theta\sigma^2/4$
with $v(0)=v_0=\theta\sigma^2/4$, i.e {\sl penetrable} particles, we
can simplify the thermodynamics of a fluid with the following
potential energy  
\bq \label{npw}
V_N=\sum_{i<j}\prod_{k=i}^{j-1}v(|x_{k}-x_{k+1}|),
\eq
for a configuration with $x_1\le x_2\le x_3\le\ldots \le x_N$.

Unfortunately in this case we cannot use the method of
separation of variables described in section \ref{sec:characteristic}
since ${\cal L}$ is time dependent.

Introducing the function $B^2(x)=-2 d\ln A(x)/dx$ one can then say
that according to Ito or Stratonovich calculus \cite{Gardiner1983} the
process defined by Eqs. (\ref{GOU1})-(\ref{GOU2}) satisfies the
following stochastic differential equation,  
\bq
\dot{\phi}(x)=-\frac{B^2(x)}{2}\phi(x)+B(x)\xi(x),
\eq
where $\xi(x)$ is Gaussian white noise with $\zeta=1$. The
$\xi(x)$ can be generated on a computer as pseudo random numbers on a
large interval $\xi\in[-a,a]$ with $a$ big enough.

%%%%%%%%%%%%%%%%%%%%%%%%%%%%%%%%%%%%%%%%%%%%%%%%%%%%%%%%%%%%%%%%%%%%%%%%%%%%%%
\subsection{Example: A non pairwise-additive Gaussian model}
%%%%%%%%%%%%%%%%%%%%%%%%%%%%%%%%%%%%%%%%%%%%%%%%%%%%%%%%%%%%%%%%%%%%%%%%%%%%%%
\label{sec:example}

For example we want to simplify the non pairwise-additive interaction
model fluid of the previous section, Eq. (\ref{npw}), with
$v(x)=v_0e^{-\gamma x^2}$, $\gamma>0$, a {\sl Gaussian core
model}. In this case we have $A(x)=e^{-\gamma x^2}$ and $B^2(x)=4\gamma
x$. For this model we expect that the attractive,
$\sigma^2=4v_0/\theta<0$, case is thermodynamically unstable in
agreement with the fact that the particles will tend to collapse at a
same point since the system is not H-stable in the sense of Ruelle
\cite{Ruelle}. On the other hand we do not know anything yet about the
repulsive, $\sigma^2>0$, case, which is H-stable and therefore we must
have $P/\theta<z$. For example, we know that there cannot be any
condensation but an interesting question is whether there can be
freezing \cite{Fantoni10b}. 

The problem (\ref{kac theorem}) becomes 
\bq \nonumber
\frac{\partial}{\partial x}\tilde{Q}(\phi,x)&=&
\left[2\gamma x\left(\sigma^2\frac{\partial^2}{\partial\phi^2}+
\frac{\partial}{\partial\phi}\phi\right)+ze^{i\phi}\right]\times\\ \label{gcm}
&&\tilde{Q}(\phi,x),\\
\tilde{Q}(\phi,0)&=&\theta_3\left(i\pi\phi/\sigma^2,e^{-2\pi^2/\sigma^2}\right)
\frac{e^{-\frac{\phi^2}{2\sigma^2}}}{\sqrt{2\pi\sigma^2}}, 
\eq
with $\tilde{Q}(\phi+2\pi,x)=\tilde{Q}(\phi,x)$. This is a
non-separable partial differential equation. Again the grand 
canonical partition function is given by Eq. (\ref{gcpf}),
\bq \nonumber
\Omega=\int_{-\pi}^\pi d\phi\,\tilde{Q}(\phi,L).
\eq

Clearly, approximating $F(\phi)\approx z$ or,
equivalently, setting $\sigma\to 0$, we get the
ideal gas behavior. In fact the solution to Eq. (\ref{gcm}) is, in this
simple case, $Q(\phi,x|0,0)=P(\phi,x|0,0)e^{zx}$, since $\partial
P/\partial x={\cal L}P$. So that from Eqs. (\ref{gcpf0}) and
(\ref{normalization}) immediately follows $\Omega=e^{zL}$.  

\begin{widetext} 
In order to make some progress towards the solution of the full
Eq. (\ref{gcm}) we define ${\cal L}\equiv x{\cal R}$ and ${\cal
  H}(x)=x{\cal R}+F$. Since ${\cal H}$ at different times do not
commute we use the following Dyson series  
\bq 
Q(\phi,x|0,0)&=&{\cal U}(x,0)R(\phi,0),\\ \nonumber
{\cal U}(x,x_0)&=&1+
\sum_{n=1}^\infty\int_{x_0}^xdx_n\int_{x_0}^{x_n}dx_{n-1}
\cdots\int_{x_0}^{x_2}dx_1\,{\cal H}(x_n)\cdots{\cal H}(x_1),
\eq
Where $R$ is given by Eq. (\ref{GOU1}). So that we find
$\Omega=1+\sum_{n=1}^\infty c_n$ with 
\bq
c_n&=&\int_{0}^{L}dx_n\int_{0}^{x_n}dx_{n-1}
\cdots\int_{0}^{x_2}dx_1\int_{-\infty}^{\infty}d\phi\,
{\cal H}(x_n)\cdots{\cal H}(x_1)R(\phi,0). 
\eq

Solving for $c_n$ we easily find $c_n=\sum_{k=1}^na_{n,k}$ with
\bq
a_{n,k}&=&\frac{e^{-k^2\sigma^2/2}f_{n,k}(\sigma^2)\gamma^{n-k}L^{2n-k}z^k}{k!},
\eq
with $f_{n,n}=1$, $f_{n,1}=0$ for $n>1$ and $f_{n,k}(\psi)$ a
polynomial of degree $n-k$ in $\psi$ beginning with the monomial of
degree one and the others of alternating signs. So
\bq \nonumber
\Omega&=&1+\sum_{n=1}^\infty\sum_{k=1}^na_{n,k}=
1+\sum_{k=1}^\infty\sum_{n=k}^\infty a_{n,k}\\ \nonumber
&=&1+\sum_{k=1}^\infty\frac{e^{-k^2\sigma^2/2}(zL)^k}{k!} 
\sum_{n=k}^\infty f_{n,k}(\sigma^2)(\gamma L^2)^{n-k}\\
&=&1+\sum_{k=1}^\infty\frac{(zL)^k}{k!}h_k(\sigma^2,\gamma L^2)
=\Omega(zL,\sigma^2,\gamma L^2), 
\eq
where we defined
\bq
h_k(\psi,\eta)&\equiv& e^{-k^2\psi/2}g_k(\psi,\eta),\\
g_k(\psi,\eta)&\equiv& 
\sum_{m=0}^\infty f_{k+m,k}(\psi)\eta^m.
\eq

First of all notice that, if the thermodynamic limit exists, we must
have $P=O(z^2/\gamma,\sigma^2)z\theta$ with $O$ a given
function of two variables such that $\lim_{\sigma\to 0}O(a,\sigma^2)=1$. 
Note that when there is no interaction between the particles $v_0\to
0$ and/or at very high temperature $\theta\to\infty$, then
$\sigma\to 0$ and we end up with an ideal gas. 

Then, if it was $h_k=1$ we would immediately find the ideal gas
behavior. On the other hand if it was $g_k=1$ we would find
an unstable system for $v_0<0$ and a stable system with $P=0=n$ for
$v_0>0$ since 
\bq
\frac{1}{L}\ln
\left[\sum_{k=0}^\infty\frac{e^{-k^2\sigma^2/2}(zL)^k}{k!}\right]\to
\left\{\begin{array}{lll}
0      & \sigma^2 > 0 & \mbox{for $L\to\infty$}\\
\infty & \sigma^2 < 0 & \mbox{for any $L$}
\end{array}\right..
\eq 
We then need to find the true $h_k$ or $g_k$.
We already know that $g_1=1$. What can we say about $g_k(\psi,\eta)$
for $k>1$? By inspection of the first few terms of the Dyson series we
find that $g_k(\psi,\eta)=1+\sum_{m=1}^\infty f_{k+m,k}(\psi)\eta^m$ with
$f_{k+m,k}(\psi)=\sum_{i=1}^m (-1)^{m+i}d_{k+m,k,i}\,\psi^i$ and
$d_{k+m,k,i}$ some positive coefficients. So that of course
$h_k(0,\eta)=1$ for all $k$, as it should. Now we can write 
\bq \nonumber
g_k(\psi,\eta)&=&1+\sum_{m=1}^\infty\sum_{i=1}^md_{k+m,k,i}\,(-\psi)^i(-\eta)^m=
1+\sum_{i=1}^\infty(-\psi)^i\sum_{m=i}^\infty d_{k+m,k,i}(-\eta)^m\\ \label{gk}
&=&1+\sum_{i=1}^\infty l_{k,i}(\eta)(-\psi)^i,
\eq 
\end{widetext}
where we defined 
\bq \label{lki}
l_{k,i}(\eta)\equiv\sum_{m=i}^\infty d_{k+m,k,i}\,(-\eta)^m.
\eq
We start looking for the coefficients for $i=1$. By inspection of the
first seven $n$ we find, for $2\le k\le n-1$,
\bq
d_{n,k,1}&=&2^n\frac{k!}{n!}b_{n,k},\\
\frac{b_{n,k}}{b_{n,k+1}}&=&(k-1)R_{n-k+2},\\
b_{n,n-1}&=&\binom{n}{n-3}\frac{1}{2^n}.
\eq
So that 
\bq
b_{n,k}=b_{n,n-1}\frac{(n-3)!}{(k-2)!}\prod_{q=k}^{n-2}R_{n-q+2},
\eq
and
\bq
d_{n,k,1}=\frac{k(k-1)}{3!}r_{n-k},
\eq
with, for $2\le k\le n-2$,
\bq 
r_{n-k}&=&\prod_{p=4}^{n-k+2}R_p,\\ \nonumber
r_2&=&4/(2\cdot2+1)!!,\\ \nonumber
r_3&=&8\cdot3/4(2\cdot3+1)!!,\\ \nonumber
r_4&=&16\cdot3/5(2\cdot4+1)!!,\\ \nonumber
r_5&=&32\cdot3/6(2\cdot5+1)!!,
\eq
and so on. We can then guess that
\bq
r_{m}=\frac{2^m3}{(2m+1)!!(m+1)}.
\eq
Then we can easily re-sum the series of Eq. (\ref{lki}) to say that 
\bq \nonumber
l_{k,1}(\eta)&=&k(k-1)\times\\
&&\frac{\mbox{}_2F_2(\{1,1\},\{3/2,2\},-x)-1}{2},
\eq
with $\mbox{}_2F_2$ a hyper-geometric function. We also find 
$\lim_{\eta\to\infty}l_{k,1}(\eta)=-k(k-1)/2$. What about
$l_{k,i}(\eta)$ for $i>1$? 

Their determination is quite laborious but let us suppose first that
we had found for $l_{k,i}$, 
\bq
l_{k,i}(\eta)=\frac{1}{i!}\left(\frac{k^2}{2}\right)^i
\left(\frac{-\eta}{1+\eta}\right)^i.
\eq
Then it would follow
\bq
h_k(\psi,\eta)=e^{-\frac{k^2}{2}\psi}e^{\frac{k^2}{2}\frac{\psi\eta}{1+\eta}}=
e^{-\frac{k^2}{2}\frac{\psi}{1+\eta}},
\eq
and for the partition function we would find
\bq \label{previous}
\Omega_L(z)=\sum_{k=0}^\infty\frac{(zL)^k}{k!}
e^{-\frac{k^2}{2}\frac{\sigma^2}{1+\gamma L^2}}.
\eq
We could then immediately say that the attractive,
$\sigma^2<0$, case would be thermodynamically unstable
since the series in Eq. (\ref{previous}) would be not summable,
whereas the repulsive, $\sigma^2>0$, case would be stable.   
In this latter case $O=\lim_{L\to\infty}\ln\Omega_L/L$ would be finite
and the system would admit a well defined thermodynamic limit without
phase transitions. The equation of state would be
\bq
\frac{P}{\theta}&=&\lim_{L\to\infty}\frac{\ln\Omega_L(z)}{L}=
O(z/\sqrt\gamma,\sigma^2)\sqrt\gamma,\\
n&=&\lim_{L\to\infty}z\frac{\Omega_L\left(ze^{-\frac{\sigma^2}{1+\gamma
    L^2}}\right)}{\Omega_L(z)}e^{-\frac{\sigma^2}{2(1+\gamma L^2)}}=z,
\eq
so that $P=O(n^2/\gamma,v_0/\theta)n\theta$ and for small
$n$ one would have $P\approx n\theta$.

In order to make some progress towards the exact solution we can then
write $d_{n,k,i}=d_{n,k,1}E_{n,k,i}$ and note that $E_{n,k,1}=1$ and
by inspection $E_{2+i,2,i}=1$. Now if we had $E_{n,k,i}=1$ for all
$n,k,i$ then we would get  
\bq \nonumber
l_{k,i}(\eta)&=&\frac{k(k-1)2^{i-1}(-\eta)^i}{(i+1)(2i+1)!!}\times\\
&&\mbox{}_2F_2(\{1,1+i\},\{3/2+i,2+i\},-\eta).
\eq 
We can then use the following limit
\bq \nonumber
&&\lim_{\eta\to\infty}\mbox{}_2F_2(\{1,1+i\},\{3/2+i,2+i\},-\eta)\eta=\\
&&\frac{(i+1)(2i+1)}{2i},
\eq
to say that 
\bq \label{lel}
\lim_{\eta\to\infty}\frac{l_{k,i}(\eta)}{(-\eta)^{i-1}}
=-\frac{k(k-1)2^{i-2}}{i(2i-1)!!}.
\eq

Since, according to Eqs. (\ref{lel}) and (\ref{gk}), in the large
$\eta$ limit,  
\bq \nonumber
g_k(\psi,\eta)&\to&1+k(k-1)\psi\times\\
&&\mbox{}_2F_2(\{1,1\},\{3/2,2\},\psi\eta)/2,
\eq
for the repulsive, $\sigma^2>0$, system we would find
\bq \nonumber
\frac{P}{\theta}&=&\lim_{L\to\infty}\frac{\ln\left[
\mbox{}_2F_2(\{1,1\},\{3/2,2\},\sigma^2\gamma L^2)\right]}{L}\\
&=&\left\{\begin{array}{lll}
\infty         & \sigma^2\gamma          & \mbox{independent of}~~$L$\\
\alpha & \sigma^2\gamma L=\alpha & \mbox{independent of}~~$L$\\
0              & \sigma^2\gamma L^2      & \mbox{independent of}~~$L$
\end{array}\right.
\eq 
and $n=0$. So that in the first two cases we would violate the
H-stability condition according to which $P/\theta<z$. This is a
signal that our approximation is too brute.  

In the appendix we report the first few exact $E_{n,k,i}$. Even if we
found it too hard to guess the full analytic expression from the first
few of them, the results of the appendix can be used to refine our
analysis. 

Our final expression for the partition function is
\bq \nonumber
\Omega&=&
\sum_{k=0}^\infty\frac{e^{-\sigma^2k^2/2}(zL)^k}{k!}\left(1+k(k-1)
\sum_{i=1}^\infty(\sigma^2\gamma L^2)^i\times\right. \\ 
&&\sum_{l=0}^\infty(-\gamma L^2)^l
\frac{E_{k+l+i,k,i}2^{l+i}}{2(l+i+1)(2(l+i)+1)!!}\Bigg)\\ \label{hstability}
&&\left\{\begin{array}{ll}
<e^{zL}  & \sigma^2>0,\\
=\infty & \sigma^2<0,
\end{array}\right.
\eq
Note that the dependence of $E_{k+m,k,i}$ on $k$ is crucial because
otherwise we could immediately conclude that the pressure would be
independent from $z$. And this fact, added to the H-stability
condition $P/\theta<z$, would be enough to say that the repulsive
Gaussian core model only admits a zero pressure zero density state.    
Note also that the dependence of $E_{n,k,i}$ on $i$ is also crucial
because otherwise for $\sigma=1$ the argument of the first two series
would be symmetric under exchange of $i$ and $l$ which would mean that
the two models with $\gamma>0$ and with $\gamma<0$ would have the same
thermodynamics which is clearly absurd \cite{Rybicki1971}. 

The first alternating series has very slow numerical convergence
as $L$ grows. We then found it difficult to extract even a numerical
equation of state. Nonetheless we found that the triple series is
convergent at least in the high temperature regime, $0<\sigma^2\ll
1$. 

From the H-stability condition (\ref{hstability}) we find that for any
$L$ and $k>1$ we must have
\bq \nonumber
&&\frac{-1}{k(k-1)}\\ \nonumber 
&\le&\sum_{i=1}^\infty(\sigma^2\gamma L^2)^i
\sum_{l=0}^\infty(-\gamma L^2)^l
\frac{E_{k+l+i,k,i}2^{l+i}}{2(l+i+1)(2(l+i)+1)!!}\\ \nonumber
&=&\sum_{m=1}^\infty(-\gamma L^2)^m
\frac{2^m\sum_{i=1}^m(-\sigma^2)^iE_{k+m,k,i}}{2(m+1)(2m+1)!!}=
G_k(\sigma^2,\gamma L^2)\\ \label{hstab}
&<&\frac{e^{\sigma^2 k^2/2}-1}{k(k-1)}.
\eq

Then, we find $\sum_{i=1}^m(-\sigma^2)^iE_{k+m,k,i} =
-\sigma^2+F_{m,k}(\sigma^2)$ for $m\ge 2$ with
$F_{m,k}(\sigma^2)=\sum_{i=2}^{m}(-\sigma^2)^{i}E_{k+m,k,i}$. In the  
large $L$ limit we then have, for $\gamma>0,$
\bq
G_k(\sigma^2,\gamma L^2)&\to& \sigma^2/2 + \lim_{L\to\infty
}H_k(\sigma^2,\gamma L^2),\\
\nonumber
H_k(\sigma^2,\gamma L^2)&=&\sum_{m=2}^\infty(-\gamma L^2)^m
\frac{2^{m-1}F_{m,k}(\sigma^2)}{(m+1)(2m+1)!!}\\
&=&(\gamma L^2)^2M_k(\sigma^2,\gamma L^2),\\ \nonumber
M_k(\sigma^2,\gamma L^2)&=&\sum_{m=0}^\infty(-\gamma L^2)^m
\frac{2^{m+1}F_{m+2,k}(\sigma^2)}{(m+3)(2m+5)!!}.
\eq 
In view of the H-stability upper bound of Eq. (\ref{hstab}), $M_k$
should be decaying as $1/L^4$ or faster, at large $L$. If it decays
faster, then $G_k$ is independent of $k$ and the only possible state
is a zero pressure one. If it decays as $1/L^4$, from the results of
the appendix we can say that it does not increase with $k$ and again
the zero pressure state is the only one possible in the thermodynamic
limit. So, in the end, we were unable to find a regular thermodynamics
even for the repulsive stable case with positive $\gamma$. Everything
is pointing towards a zero pressure state in the thermodynamic 
limit. This would be in agreement with the observation that as 
$\theta\to 0$ the only configurations contributing to the integral in
Eq. (\ref{omega1}) are the ones with minimum $V_N-\mu N$ which are
those where the particles are infinitely spaced one another with $n\to
0$. 

%%%%%%%%%%%%%%%%%%%%%%%%%%%%%%%%%%%%%%%%%%%%%%%%%%%%%%%%%%%%%%%%%%%%%%%%%%%%%%
\subsection{Observation}
%%%%%%%%%%%%%%%%%%%%%%%%%%%%%%%%%%%%%%%%%%%%%%%%%%%%%%%%%%%%%%%%%%%%%%%%%%%%%%
\label{sec:observation}

Now, we can observe that applying the previous analysis to the
pairwise-additive Kac-Baker model, $v(x)=v_0e^{-\gamma|x|}$, $\gamma>0$, the
structure of the solution for the partition function reads  
\bq \nonumber
\Omega&=&1+\sum_{k=1}^\infty\frac{e^{-k^2\sigma^2/2}(zL)^k}{k!} 
\sum_{n=k}^\infty f_{n,k}(\sigma^2)(\gamma L)^{n-k}\\
&=&\Omega(zL,\sigma^2,\gamma L),
\eq
with some given polynomials $f_{n,k}$. Again we can definitely say
that the attractive model is thermodynamically unstable. But we know
that the repulsive case is a pairwise-additive interaction model with
a regular thermodynamics. 

%%%%%%%%%%%%%%%%%%%%%%%%%%%%%%%%%%%%%%%%%%%%%%%%%%%%%%%%%%%%%%%%%%%%%%%%%%%%%%
\section{Conclusions}
%%%%%%%%%%%%%%%%%%%%%%%%%%%%%%%%%%%%%%%%%%%%%%%%%%%%%%%%%%%%%%%%%%%%%%%%%%%%%%
\label{sec:conclusions}

We reviewed, under the unified setting of functional integration in
one dimension, some of the exactly solvable one dimensional continuum
fluid models of equilibrium classical statistical mechanics. Following
the original idea of Marc Kac we write the partition function of each
model as a path integral over particular Markoffian, Gaussian
stochastic processes. Following the idea of Sam Edwards we further
reduce the thermodynamic problem for such fluids to the solution of a
second order ordinary differential equation, the characteristic value
problem. 

In the work of Edwards and Lenard \cite{Edwards62} it is also given a
detailed analysis of how one can extend this method to get solutions
for the pair- and higher orders static correlation functions. 

We propose a generalization of the method which allows to treat other
models with a non pairwise-additive interaction between the constituent
penetrable particles of the fluid. The characteristic 
value problem of Edwards cannot be used anymore but the simplification
of Kac remains valid. We apply this further developments to the simple
case of the Gaussian core fluid model for which we prove that the
attractive system is thermodynamically unstable, in agreement with the
fact that it is not H-stable in the sense of Ruelle \cite{Ruelle}, and
find an approximate expression for the exact partition function in
terms of a triple series one of which is alternating. We were unable
to find a well defined thermodynamics even for the repulsive
system. Everything suggest that the only admitted state in the
thermodynamic limit is the zero pressure one. 

\appendix
%%%%%%%%%%%%%%%%%%%%%%%%%%%%%%%%%%%%%%%%%%%%%%%%%%%%%%%%%%%%%%%%%%%%%%%%%%%%%%
\section{The coefficients $E_{n,k,i}$} 
%%%%%%%%%%%%%%%%%%%%%%%%%%%%%%%%%%%%%%%%%%%%%%%%%%%%%%%%%%%%%%%%%%%%%%%%%%%%%%
\label{app:}

%\begin{widetext}
In table \ref{tab:e} we list the first exact $E_{n,k,i}$ coefficients
for $i=2,3,4$ and the first seven $n$. 

\begin{table*}[htb]
\caption{Exact $E_{n,k,i}$ for $i=2,3,4$.}
%\begin{ruledtabular}
\begin{tabular}{|l||l|l|l|l|}
\hline
$E_{n,k,2}$ & $n=7$ & $n=6$ & $n=5$ & $n=4$ \\   
\hline
\hline
$k=2$ & $15$ & $7$ & $3$ & $1$\\ 
\hline
$k=3$ & $5105/352$ & $79/12$ & $33/14$ & \\ 
\hline
$k=4$ & $211/18$ & $243/56$ & & \\ 
\hline
$k=5$ & $389/56$ & & & \\ 
\hline
\end{tabular}
\begin{tabular}{|l||l|l|l|}
\hline
$E_{n,k,3}$ & $n=7$ & $n=6$ & $n=5$ \\   
\hline
\hline
$k=2$ & $25$ & $6$ & $1$\\ 
\hline
$k=3$ & $4923/176$ & $31/6$ & \\ 
\hline
$k=4$ & $17$ & & \\ 
\hline
\end{tabular}
\begin{tabular}{|l||l|l|l|l|}
\hline
$E_{n,k,4}$ & $n=7$ & $n=6$ \\   
\hline
\hline
$k=2$ & $10$ & $1$ \\ 
\hline
$k=3$ & $965/88$ & \\ 
\hline
\end{tabular}
%\end{ruledtabular}
\\
%\begin{ruledtabular}
\begin{tabular}{|l||l|l|l|l|}
\hline
$E_{n,k,2}$ & $n=7$ & $n=6$ & $n=5$ & $n=4$ \\   
\hline
\hline
$k=2$ & $15$ & $7$ & $3$ & $1$\\ 
\hline
$k=3$ & $14.5$ & $6.6$ & $2.4$ & \\ 
\hline
$k=4$ & $11.7$ & $4.3$ & & \\ 
\hline
$k=5$ & $6.9$ & & & \\ 
\hline
\end{tabular}
\begin{tabular}{|l||l|l|l|}
\hline
$E_{n,k,3}$ & $n=7$ & $n=6$ & $n=5$ \\   
\hline
\hline
$k=2$ & $25$ & $6$ & $1$\\ 
\hline
$k=3$ & $28.0$ & $5.2$ & \\ 
\hline
$k=4$ & $17$ & & \\ 
\hline
\end{tabular}
\begin{tabular}{|l||l|l|l|l|}
\hline
$E_{n,k,4}$ & $n=7$ & $n=6$ \\   
\hline
\hline
$k=2$ & $10$ & $1$ \\ 
\hline
$k=3$ & $11.0$ & \\ 
\hline
\end{tabular}
%\end{ruledtabular}
\label{tab:e}
\end{table*}

From the table we can see how there is a very weak dependence on
$k$. So we can on a first ground assume that $E_{n,k,i}\approx
E_{n,2,i}=e_{n-i,i}$ for all $k$. Moreover the entries of the table
satisfy the following recurrence relation 
\bq
e_{2,i}&=&1,\\
e_{j,2}&=&2^{j-1}-1,\\
e_{j,i}&=&ie_{j-1,i}+e_{j,i-1},
\eq
with $j=n-i$. So that introducing the generating function
$\varphi(x,i)=\sum_{j=2}^\infty e_{j,i}x^j$ we easily find
\bq
\varphi(x,2)&=&x^2/(x-1)(2x-1),\\
\varphi(x,i)/x&=&i\varphi(x,i)+\varphi(x,i-1)/x,
\eq
with solution
\bq
\varphi(x,i)=\frac{x^3}{(x-1)(-x)^i(2-1/x)_{i-1}},
\eq
with $(a)_i=a(a+1)\cdots(a+i-1)=\Gamma(a+i)/\Gamma(a)$ the Pochhammer
symbol. The desired coefficient $e_{j,i}$ is the $j$-th coefficient in
the series expansion of $\varphi(x,i)$ around $x=0$.

More precisely we can then write $E_{n,k,i}=h_{n,k,i}e_{n-i,i}$
with $h_{n,2,i}=1$ and $E_{2+i,2,i}=1$. We can also observe that
$E_{n,k,i}$ tends to decrease with $k$ at fixed $n$ and $i$.

%\end{widetext}
%%%%%%%%%%%%%%%%%%%%%%%%%%%%%%%%%%%%%%%%%%%%%%%%%%%%%%%%%%%%%%%%%%%%%%%%%%%%%% 
\begin{acknowledgments}
I would like to thank Prof. Klaus Shulten whose course in
non-equilibrium statistical mechanics held at Urbana in 1999
in the Loomis laboratory, stimulated the study of the EL paper as a
final individual project. I would also like to thank Prof. Giorgio
Pastore and Andres Santos for many useful discussions. 
\end{acknowledgments} 
%%%%%%%%%%%%%%%%%%%%%%%%%%%%%%%%%%%%%%%%%%%%%%%%%%%%%%%%%%%%%%%%%%%%%%%%%%%%%%
%\bibliographystyle{}
\bibliography{1d}
%%%%%%%%%%%%%%%%%%%%%%%%%%%%%%%%%%%%%%%%%%%%%%%%%%%%%%%%%%%%%%%%%%%%%%%%%%%%%%
%%%%%%%%%%%%%%%%%%%%%%%%%%%%%%%%%%%%%%%%%%%%%%%%%%%%%%%%%%%%%%%%%%%%%%%%%%%%%%
%%%%%%%%%%%%%%%%%%%%%%%%%%%%%%%%%%%%%%%%%%%%%%%%%%%%%%%%%%%%%%%%%%%%%%%%%%%%%%
\end{document}